\documentclass[conference]{IEEEtran}
\ifCLASSINFOpdf
\else
\fi
\usepackage{hyperref}
\usepackage{graphics} 
\usepackage{url}      
\usepackage{comment}
\usepackage{multirow}
\usepackage{epstopdf}
\usepackage{graphicx}
\usepackage{caption}
\usepackage{subcaption}
\usepackage{algorithm}
\usepackage{algpseudocode}

\floatname{algorithm}{Procedure}

\begin{document}
%
\title{A Relevant Content Filtering Based Framework For Data Stream Summarization}

\author{\IEEEauthorblockN{Cailing Dong}
\IEEEauthorblockA{Department of Information Systems\\
University of Maryland, Baltimore County\\
Baltimore, Maryland 21250\\
Email: cailing.dong@umbc.edu}
\and
\IEEEauthorblockN{Arvind Agarwal﻿﻿}
\IEEEauthorblockA{Palo Alto Research Center (PARC) \\
Webster, New York 14580\\
Email: arvind.agarwal@xerox.com}
}


%


\maketitle

\begin{abstract}
Social media platforms are a rich source of information these days, however, of all the available information, only a small fraction is of users' interest. To help users catch up with the latest topics of their interests from the large amount of information available in social media, we present a relevant content filtering based framework for data stream summarization. More specifically, given the topic or event of interest, this framework can dynamically discover and filter out relevant information from irrelevant information in the stream of text provided by social media platforms. It then further captures the most representative and up-to-date information to generate a sequential summary or event story line along with the evolution of the topic or event. Our framework does not depend on any labeled data, it instead uses the weak supervision provided by the user, which matches the real scenarios of users searching for information about an ongoing event. We experimented on two real events traced by a Twitter dataset from TREC 2011. The results verified the effectiveness of relevant content filtering and sequential summary generation of the proposed framework. It also shows its robustness of using the most easy-to-obtain weak supervision, i.e., trending topic or hashtag. Thus, this framework can be easily integrated into social media platforms such as Twitter to generate sequential summaries for the events of interest. We also make the manually generated gold-standard sequential summaries of the two test events publicly available\footnote{\label{note1}\scriptsize{\url{https://drive.google.com/open?id=15jRw13i0xARUW3HqBn3BdR45IXk7P2Qj-HO__OFmMW0}}} for future use in the community. 

\end{abstract}


%
\IEEEpeerreviewmaketitle

\section{Introduction}
In the last few years,  social media, in particular micro-blogging websites, has seen a steep rise in their popularity with the increasing number of users contributing the content in terms of short text messages.  As of the 4th quarter 2015, there are about 305 million monthly active users on Twitter who posts more than 500 million Tweets everyday. When so much information coming at such a high speed, it is of vital importance to provide a concise and up-to-date summary to help users catching up and understanding the topic and event of interests. Besides the large quantity of the posts, these short text messages are status updates consists of various and ever-changing topics, ranging from simple status updates about personal life such as {\it going to visit a friend}, to text messages about ongoing real-life events such as {\it FIFA world cup}, {\it Justin Bieber performance}, to more involved conversations about the topics of general interests such as {\it global warming}, {\it terrorism} etc. Although some of these posts are related to events of general interests, most are simply about daily-life routine events, and therefore, of little-to-no interest to a user other than the one who posted them, or to his immediate connections. In fact, a  study shows that among all the posts on Twitter, only 3.6\% of the posts are related to the topics of mainstream news~\cite{TwitterStudy2009}. Therefore, there is an apparent need for relevant content filtering, especially when the information is coming at such a high speed and with so much noise. 

The usefulness of relevant content filtering system goes beyond the users searching for information on the website. This kind of system is useful for many end-user applications, especially if they were to operate on streaming data, such as sentiment analysis \cite{agarwal2011sentiment}, summarization \cite{tweetmotif2010}, topic detection\cite{diao2012finding}, etc. Almost in all of the tasks that operate on Twitter data, the very first step should be removing uninteresting information. However, most of the work on micro-blogging summarization has not put as much effort on relevant content filtering before performing summarization. One major reason is the streaming nature of the data makes it hard for one to rely on a {\it static} method for information filtering. One instead needs a method that is able to filter content dynamically to track the evolving news and events.

In this paper, we present a relevant content filtering based
framework for data stream summarization, namely \textbf{W}eakly \textbf{S}upervised \textbf{S}tream \textbf{F}ilter and \textbf{S}ummarizer (W$S^2$FS), which is suitable for both \emph{relevant content filtering} and {\it sequential summarization}. Unlike classical supervised method that relies on the availability of labeled data, the proposed framework does {\it not} use any manually labeled data, it instead uses \textit{weak supervision} from users, which can be as simple as topical keywords, or in the form of any rule that can provide global {\it feature-level} information. When using the most easy-to-obtain supervision, i.e., hashtags, our framework can be treated as an almost unsupervised method, making it practical to be used for summarizing both {\it personalized events} (i.e. events of personal interests) and {\it general events} (events of general interests). Another important strength of the framework is its ability to handle streaming data. The framework is modeled as an online classification framework,  which evolves i.e., learns from the new data as it becomes available. Its independence on any labeled data and its ability to adapt to streaming data make it suitable for integration into streaming data websites such as Twitter, where it can generate up-to-date summaries for topics/events of interests. In general, our contribution in this paper is as follows:
\begin{itemize}
\item We propose a relevant content filtering based framework for data stream summarization. It couples the two important tasks in social media, i.e., {\it relevant content filtering} and {\it event summarization} in one integrated framework. This framework is not only able to capture the event evolution and dynamically filter out relevant content, but also generates sequential summary or event story line effectively. 
\item The proposed framework is almost unsupervised since it does not use any manual labeled data\footnote{\scriptsize{It is not explicitly labeled for the classification task, rather than obtained from the data itself.}}. The {\it weak supervision} it uses can either be done automatically (making it fully unsupervised) or be provided by information seeker.
\item It best simulates the real scenarios of users searching for relevant information with self-defined search queries from any social stream websites, hoping to get a concise and informative summary. Thus, it can be integrated into any social stream websites such as Twitter readily to generate the summaries of the event of interest in a timely fashion.
\item Our experimental results showed that the hashtag-based {\it weak supervision} produces the best results. As such supervision is easy to obtain, our framework can be easily extended to generate both personal event summary and global event summary.
\item We make the manually generated chunk-wise and sequential summaries of two real test events publicly available, which can be used readily in the community.
\end{itemize}

The remainder of the paper is structured as follows. Section~\ref{sec:RelatedWork} provides some related work. In Section~\ref{sec:framework}, we describe the general structure, major components and detailed implementation of the proposed framework. Section~\ref{sec:exp} demonstrates the comprehensive experiments on two real events delivered in Twitter, and examines its performance using different types of weak supervisions. Finally we discuss and conclude the paper in Section~\ref{sec:discussion} and Section~\ref{sec:conclusion}. 

\section{Related Work}\label{sec:RelatedWork}

\subsection{Relevant Content Filtering}
Most of the the work in content filtering has been based on the following two types of methods: (1) information retrieval based methods, (2) machine learning based method. In information retrieval based methods, a query is formed based on the information that is being sought, and then, the query is executed to find the relevant content. Although in theory, any traditional information retrieval based method can be used for this, streaming nature of the data on micro-blogging website makes it hard to implement. For online streaming social media content, using information retrieval method that employs pre-built indexes is not feasible. Although a mature field in itself, information retrieval field has not yet found its ground in retrieving the content from streaming social media platforms. Most of the current work still relies on simple method such as query keywords based search. In machine learning based methods, a typical approach is to build topic specific {\it supervised} classifier \cite{khan2013improved}. However, these supervised classification methods have various limitations which makes them less appropriate for content filtering for streaming data. First of all, supervised classification methods need labeled data. Getting labeled data is both expensive and time consuming. Secondly, supervised classification methods are not easily extensible to new topics. Every time a new topic comes, one has to create new labeled data and then build a new classifier. Since the topics keep evolving in the data stream, it is not reliable to use a fixed labeled dataset to capture the whole event. 

\subsection{Micro-blogging Summarization}
Previous work on micro-blogging summarization can be divided into three categories, i.e., frequency-based methods~\cite{sharifi2010experiments,olariu2013hierarchical}, graph-based methods~\cite{sharifi2010summarizing,olariu2014efficient} and context-based methods~\cite{chang2013towards,yang2012framework}.
Frequency-based methods are based on the assumption that if a word or a set of words in a data instance (such as a tweet) has a high frequency of being repeated, the instances containing the set of high-frequency words must be good candidates for generating summary. Based on the similar assumption, graph-based methods build a word graph to capture common sequences of words about the given topic. The path with the highest total weight is regarded as a candidate summary instance. Typical graph-based methods include TextRank~\cite{textrank2004}, LexRank~\cite{erkan2004lexrank} and Phrase Reinforcement (PR) algorithm~\cite{sharifi2010summarizing}. Context-based approaches rate the importance of a data instance not only based on its textual importance, but also based upon other non-textual features, such as user influence, data instance popularity and temporal signals~\cite{chang2013towards}. Although verified to be very effective in generating single-sentence summary, none of these algorithms were specifically designed for or have been used on streaming data. Furthermore, these methods are pure summarization methods assuming the relevant content is ready to use. Putting them into the streaming environment, the effectiveness of these summarization methods would not be guaranteed as they could fail to capture the evolution of the given topic based on {\it static} keywords. 

In contrast, our proposed framework integrates both {\it relevant content filtering} and {\it summarization}. It is dynamic and changes its behavior according to the arriving data from the stream. We emphasize here that these summarization methods are not competitors to the proposed framework, they are rather complementary, i.e., any of these summarization methods can be integrated with the {\it relevant content filter} of our framework.

\subsection{Event Tracking and Summarization}
Lately, event tracking and summarization has raised lots of attention, where one key task is to detect the relevant content about the event. One major application domain is summarizing scheduled events~\cite{chakrabarti2011event,nichols2012summarizing}. For instance, Chakrabarti et al.~\cite{chakrabarti2011event} employed a modified Hidden Markov Model (HMM) to learn the structure and vocabulary of multiple American football games, in order to detect relevant content with regard to future games and further summarize them. Nichols et al~\cite{nichols2012summarizing} used an unsupervised algorithms to generate summary for sporting events, in which relevant content were extracted by detecting spikes in volume of status updates and further ranking them using a phrase graph. Using the similar approach, Zubiaga achieved real-time summarization of scheduled sporting events~\cite{zubiaga2012towards}. Compared with scheduled events which usually have specific ``moments'' and terminology, tracking unscheduled events are more challenging, but of general applicability. In~\cite{osborne2014real}, Osborne et al. classified a tweet as relevant or not based on the score distribution within a set of tweets, and further generated summary by removing redundancy among the selected tweets. To build a large-scale corpus for evaluating event detection on Twitter, McMinn et al.~\cite{mcminn2013building} employed crowdsourcing to gather relevance judgements. Other work mostly focus on detecting data volume changes, extracting sub-topics and further clustering them into the same events~\cite{long2011towards,marcus2011twitinfo}. In contrast to these methods, our framework employs more sophisticated techniques grounded in machine learning (i.e. online learning) for filtering out relevant content from irrelevant one, and for summarizing events. Furthermore, our framework is designed for all events, scheduled or unscheduled.

\section{\underline{W}eakly \underline{S}upervised \underline{S}tream \underline{F}ilter and \underline{S}ummarizer (WS$^{2}$FS)}
\label{sec:framework}

To filter out the relevant content with regard to a given topic or event from a data stream, we need to classify each instance (e.g., each tweet) into ``relevant'' or ``non-relevant" categories, which is a classic binary classification problem. A good classifier for streaming data needs:\begin{itemize}
\item \textit{Reliable training datasets.} For a text stream containing almost infinite set of topics, creating such training datasets through manual annotations is impractical. It calls for an automatic approach to creating reliable training datasets.
\item \textit{Maintain the ``main thread'' and capture the ``evolution'' of the event}. A good classifier for streaming data should be continuously learning. It should capture not only the main ``theme'' of the event, but also the content as the event unfolds, which is also an important feature of building a good summarizer on streaming data. 
\end{itemize}

\begin{figure}[h]
\centering
\includegraphics[width=0.9\linewidth]{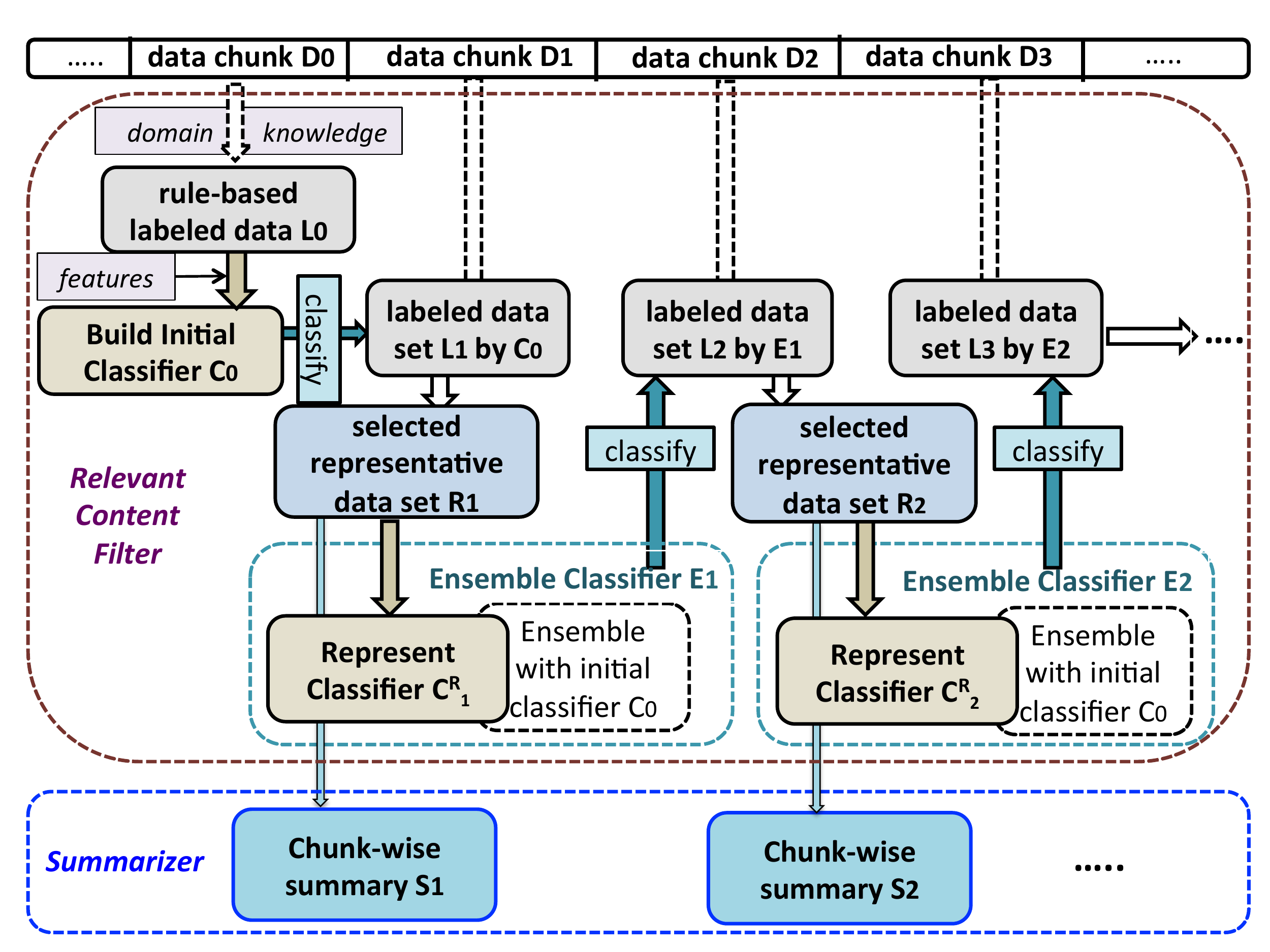}
\caption{Framework of \footnotesize{WS$^{2}$FS}}\label{fig:framework}
\end{figure}

Motivated by the above two important tasks, we propose a \textbf{W}eakly \textbf{S}upervised \textbf{S}tream \textbf{F}ilter and \textbf{S}summarizer (\textbf{WS$^{2}$FS}) to filter out relevant content and further generate sequential summary from data stream.  
The general framework is shown in Figure~\ref{fig:framework}. 

\subsection{Relevant Content Filter of WS$^2$FS} \label{subset:general}
The general structure of the {\it relevant content filter} in WS$^2$FS is demonstrated at the top part of Figure~\ref{fig:framework}. The first step is to define the appropriate size of a stream chunk and thereafter split the data stream into different data stream chunks. The size of the chunk represents the granularity of filtering. It is flexible to define, e.g., according to timestamp, data volume, etc.  The next step is choosing an appropriate starting point. Although it can start from any time or any stream chunk, a good starting point is when the given event starts to emerge (similar to the spike in data volume), which is a good timing to capture the ``main thread'' of the event. In general, {\it relevant content filter} consists of three main components: (1) a one-time \textsc{Initial Classifier $C_0$} builder which builds $C_0$ using the first stream chunk $D_{0}$, (2) a representative dataset $R_{i}$ builder and (3) an \textsc{Ensemble Classifier $E_{i}$} builder, for each chunk $D_{i}$ ($i \ge 1$). In the following, we describe each of these components in detail.  

\emph{First of all, build the initial classifier $C_0$.}
Once the stream is split into chunks, an initial classifier $C_0$ is built on first chunk $D_0$. Since $D_0$ does not have labels, the very first task is to get labels for $D_0$. One major contribution of the proposed framework is that it does not use any manually labeled data. Instead, the labels of instances in $D_0$ are created automatically based on \emph{weak supervision} provided by the information seeker. We emphasize here that the labels are not obtained in a classical way, i.e., by asking label for each instance, rather, information seekers provide rules that operate on the whole corpora which in turn produce each label. This component results in a rule-based labeled dataset $L_{0}$, which is used to construct the \textsc{initial classifier} $C_{0}$.

\emph{Second and third} components correspond to building representative training datasets and classifiers for each of the following chunks. These two components function alternatively, i.e., classifier in chunk $i$ i.e., $C_i$ is used to classify the data in the next chunk i.e., $D_{i+1}$ (in other words get $L_{i+1}$) which is then further processed to build the classifier for chunk $i$+$1$, i.e., $C_{i+1}$. The training dataset for {\it each} chunk should be such that it captures the dynamic nature of the event in that chunk. In other words, it should contain any new content that appeared in that chunk. 
We build training dataset for chunk $i+1$ using classifier from chunk $i$, followed by some further processing. This further processing consists of using {\it weak supervision} along with other available information in the dataset such as the number of followers in Twitter (detailed information will be elaborated in the following part). We call such dataset as {\it chunk representative dataset} and denote it by $R_i$. The positive instances in $R_{i}$ are selected in a way that they are highly reliable, and contain subtopics of the given event in chunk ${i}$. Thus they capture the ``evolution'' of the event and set up the up-to-date criteria for filtering relevant data from the next chunk. 

In order to build a content filtering classifier for chunk $i$, we first build a classifier called \textsc{Represent Classifier} $C^R_{i}$ upon each $R_i$. As this classifier is built upon $R_i$, which mainly captures the localized subtopics that appeared in chunk $i$, it will fail to capture the main theme of the event. While, the main theme of the event is often captured by the first classifier $C_0$. In order to capture both the main theme of the event and the updated subtopics, for each chunk $i$, an ensemble classifier is built using two base classifiers, i.e., chunk-specific \textsc{Represent Classifier} $C^R_{i}$ and the \textsc{Initial classifier} $C_0$. We call this classifier \textsc{Ensemble Classifier} $E_{i}$. For each instance, its confidence of being a relevant instance is calculated by combining the confidence values produced by both of the two base classifiers, as given by:

{\footnotesize
\begin{equation}\label{eq:alpha}
Conf_{E_i}(x) = (1-\alpha) * Conf_{C_{0}}(x) + \alpha * Conf_{C^{R}_{i}}(x).
\end{equation}
} 
The weight $\alpha$ is flexible to set. If the provided {\it weak supervision} is not strong, we may need to put more weight on $C_{0}$ to better maintain the main ``theme''. On the contrary, if we are more interested in the evolution of the given event, we can put more weight on $C^{R}_{i}$ to better capture the sub-topics. 
\subsubsection {Chunk representative dataset $R_i$ builder}
How to create {\it chunk representative dataset} $R_i$ to build \textsc{Represent Classifier} $C^{R}_i$ is the key component of {\it relevant content filter } in WS$^{2}$FS. As mentioned before, the positive instances $R_i^+$  in $R_i$ should capture the evolution of the given event. This evolution is usually captured by the words that most frequently co-occurred with the provided \emph{weak supervision} or its induced rules. We call such words as \textit{companion words}. It is worth noting that the purpose of the {\it weak supervision} goes beyond getting initial labels for $D_0$. They are used all along the text stream to obtain  \textit{companion} words. 

\begin{algorithm}[htb]
\footnotesize
\caption{\small{Qualification Criteria $\mathcal{Q}$ of Selecting $R_i$}} \label{alg:criteria}
\begin{algorithmic}[1]
\Require Labeled data $L_{i}$ from $D_{i}$ and corresponding confidence value $Conf_{E_i}$ ($i \ge 0$)
\Require \#candidate instances = $p$; \#companion words = $m$; \#representative instances = $n$
\Ensure {\it chunk representative dataset} $R_{i}$ sorted by relevancy 
\item[]
\Comment \emph{\bf Select companion words}
\State \emph{$Avg^{+}(Conf_{E_i})$}: average {\it confidence} value of positive instances $L^{+}_{i}$  
\State $L^{+'}_{i} \gets \{x \in L^{+}_{i}: Conf_{E_i}(x) \ge Avg^{+}(Conf_{E_i}(x)) \}$ 
\State $L^{C+}_{i}\gets$ sort $L^{+'}_{i}$ by $correlation$ descendingly
\State $L^{R+}_{i}\gets$ sort $L^{+'}_{i}$ by $credibility$ descendingly
\State $L^{cand+}_{i}\gets$ (top-$p$ ($L^{C+}_{i}$) $\cap$ top-$p$ ($L^{R+}_{i}$)) $\cup$ [top ($L^{C+}_{i}$)]  

\State $Comp_{i}\gets$ top-$m$ most frequent words in $L^{cand+}_{i}$ 
\item[]
\Comment \emph{{\bf Select distant words}}
\State \emph{$Avg^{-}(Conf_{E_i})$}: average {\it confidence} value of negative instances $L^{-}_{i}$
\State $L^{-'}_{i}\gets  \{x \in L^{-}_{i}: (Conf_{E_i}(x) \le Avg^{-}(Conf_{E_i}(x))\}$
\State $L^{C-}_{i}\gets$ sort $L^{-'}_{i}$ by $correlation$ ascendingly
\State $L^{R-}_{i}\gets$ sort $L^{-'}_{i}$ by $credibility$ ascendingly
\State $L^{cand-}_{i}\gets$ (top-$p$ ($L^{C-}_{i}$) $\cap$ top-$p$ ($L^{R-}_{i}$)) $\cup$ [top ($L^{C-}_{i}$)] 
\State $Distant_{i}\gets$ top-$m$ most frequent words in $L^{cand-}_{i}$ 
\item[]
\Comment \emph{\bf Select representative instances and build $R_{i}$ }
\State $Common_{i} = Comp_{i} \cap Distant_{i}$ \Comment  \emph{{\bf check diversity}}
\If {$(\frac{|Common_{i}|}{|Comp_{i}|} < 0.5) \& (\frac{|Common_{i}|}{|Distant_{i}|} < 0.5)$}  
    \State $Comp_{i} \gets Comp_{i} \cup T$
    \State $S^{+}_{i} \gets$ top-$n (L^{+'}_{i})$ on frequency of $Comp_{i}$ 
    \State $S^{-}_{i} \gets$ top-$n (L^{-'}_{i})$ on frequency of $Distant_{i}$ 
    \State $R_{i} \gets S^{+}_{i} \cup S^{-}_{i}$
\EndIf
\end{algorithmic}
\end{algorithm}

The $\mathbf{qualification \: criteria}$  $\mathcal{Q}$ for selecting representative dataset $R_i$ is given by the pseudocode in Procedure~\ref{alg:criteria}. The selection of \emph{companion words} (line 1-6) depends on the ``3C'' factors, i.e., \emph{Confidence}, \emph{Correlation} and \emph{Credibility}. \emph{Confidence} means the confidence of being relevant as judged by the \textsc{Ensemble Classifier} $E_i$.  \emph{Correlation} measures the semantic relatedness to the given event. And \emph{credibility} measures the reliability of the ``source'' of the instance. 
Specifically, line 1-2 make sure the candidate positive instances have high \emph{confidence}. Line 3 and line 4 create two sorted list of instances according to \emph{correlation} and \emph{credibility} in descending order. The qualified instances are selected as the intersection of the top-$p$ instances in the two sorted lists, guaranteeing both high \emph{correlation} and \emph{credibility}. If the number of qualified instances is smaller than $p$, the top instances in the sorted list based on \emph{correlation} are added as supplements (line 5). Finally, the top-$m$ most frequent words in the qualified instances are selected as \emph{companion words} (line 6). In order to build \textsc{Represent Classifier}, we also need to select representative negative instances. The procedure is described in line 7-12 in Procedure~\ref{alg:criteria}, which is based on low values of \textit{confidence},  \textit{correlation}, and  \textit{credibility}. We call the words selected from the these negative instances as \textit{distant} words. Finally, the instances that have highest frequency of \emph{companion words} and \emph{distant words} are selected as representative positive and negative instances, respectively (line 16-17). Before doing this, we need to check the \emph{diversity} of the two sets of words (line 14). If the ratio of the size of the common words to the size of either set is low, it means the two sets of words have large diversity and therefore are reliable enough to be selected as representative instances. Otherwise, the representative dataset building process will be skipped from this given stream chunk.

\subsection{Summarizer of WS$^{2}$FS}\label{sec:summarization}
The {\it summarizer} of WS$^{2}$FS is seen at the bottom of Figure~\ref{fig:framework}. For each chunk, the {\it relevant content filter} in WS$^{2}$FS has produced a list of representative positive instances $R_i^+$, sorted based on \emph{confidence}, \emph{correlation} and \emph{credibility}. The final ranking signifies the importance and representativeness of the instances. Thus, the top-ranked instances in each stream chunk can be regarded as good candidates to generate the {\it chunk-wise summary} of the given event. Later, the {\it summarizer} of WS$^{2}$FS combines all the chunk-wise summaries in chronological order, and further process it to generate the final {\it sequential summary}. 

A key point here is how to select a candidate data instance as a final chunk-wise or sequential summary instance. As the resultant summary should cover as many aspects of the event as possible, we use {\it diversity} as the selection measurement.  {\it Diversity} refers to the opposite of redundancy here. That is, all the instances should minimally overlap with each other within the final {\it chunk-wise summary} and {\it sequential summary}. In our work, we employ ROUGE-L~\cite{lin2004rouge} to calculate the degree of overlapping between a candidate instance and each of the already chosen summary instances. ROUGE-L calculates the statistics (average recall, precision and F1 values) about the longest common subsequence between two string. Obviously, the higher the value is, the less diversity the candidate instance will bring. A threshold $\theta_{diversity}$ can be set flexibly based on the desired level of diversity. In our work, to keep higher diversity, we regard a candidate instance to be a final summary instance when all the three statistic values are less than 0.4.

\section{Experiments}
\label{sec:exp}
Twitter is a representative source of data stream.
In our experiment, we apply WS$^{2}$FS on a Tweet stream dataset \textsc{Tweets2011} from TREC 2011 Microblog Track\footnote{\scriptsize{\url{http://trec.nist.gov/data/tweets/}}}. The dataset contains 16 million tweets sampled between January 23rd and February 8th, 2011. It also provides a set of manually labeled relevant tweets for 50 topics. 
Two important events happened during the two weeks of sampled tweets, {\it Moscow airport bombing} and {\it Egyptian revolution}. In this dataset, the first event is descried by one topic {\it Moscow airport bombing}, while the second event covers three topics, {\it Egyptian curfew}, {\it Egyptian evacuation} and {\it Egyptian protesters attack museum}. We combine all the tweets related to the three topics and associate them with event {\it Egyptian revolution}. As the original annotation is based on both the tweet text and the content of the URL~\cite{Ounis2011}, we asked two annotators to re-annotate the test dataset only based on tweet text. The inter-annotator agreement is 0.952 measured by Cohen's kappa coefficient.

\subsection{Experimental Settings}

\noindent{\it (1) Stream chunks.} We split the data stream into different chunks by the creation ``date'' of the tweets.

\noindent{\it (2) Weak supervision.} In most cases, the initial knowledge of the users about an event is a general concept. Therefore, we simply define the {\it weak supervision} as: \emph{``A tweet is a relevant instance if any topical words (case insensitive) appear in the tweet"}. We also treat the results of applying the \emph{weak supervision} on the testing dataset as our \emph{Baseline} method, which is also a simulation of relevant content filtering via Twitter Search API.

 In practice, different users may provide different types/levels of {\it weak supervision} about the event. As event {\it Egyptian revolution} covers three aspects in the original datasets, we simulate the scenarios of users having the following three types of {\it weak supervision} about this event:
\begin{itemize}
\item {\it Type-a}: {\it general concept} -- when a user cares about the event in general, described by topical words ``Egyptian, revolution''; 
\item {\it Type-b}: {\it trending topic/ hashtag} -- when a user is interested in the trending topic or hashtag ``\#Jan25'';
\item {\it Type-c}: {\it specific aspects} -- when a user is interested in specific aspects of the event, described by the keywords ``Egyptian, protesters, attack, museum, curfew, evacuation'' (these keywords are chosen based on the topical words in the three topics in the original dataset). 
\end{itemize}

\noindent{\it (4) Features}. WS$^{2}$FS is designed as a general framework that does not rely on any specific features, thus we only use the following three type of features:
\begin{itemize}
\item {\it Keywords-feq}: the total frequency of topical key words in the tweet text. It is used to measure one of the ``3C'' factors, i.e., \emph{correlation}.
\item {\it Status}: We define status as the normalized ratio between the number of followers and followings of the user who wrote or retweeted the current tweet. The $status$ shows the reputation of the tweet author to some extent. Intuitively, the higher the value of $status$, the more reliable is the tweet. We use it to measure the {\it credibility} in qualification criteria $\mathcal{Q}$.
\item {\it Content-words}: the words in tweet text that carry the content. We use Twitter NLP toolkit\footnote{\scriptsize\url{http://www.ark.cs.cmu.edu/TweetNLP/}}~\cite{Gimpel2011} to get part-of-speech (POS) tag for each word in tweet text. Only the content words with specific POS tags (``N'', ``\char`\^'' , ``S'', ``Z'', ``V'', ``A'', ``R'' , ``D'' ) are kept. 
\end{itemize}

\noindent{\it (5) Starting point.} As described in Section~\ref{subset:general}, a good and meaningful starting point in the stream is when the given event starts to emerge. Thus, we choose Jan 23rd, 2011 and Jan 25th, 2011 as the starting point to collect the first stream chunk for events {\it Moscow airport bombing} and {\it Egyptian revolution}, respectively. 

\noindent{\it (6) Classifier.} As \emph{Na\"{\i}ve Bayes} has been verified to be very effective in many text mining related tasks, we use it to build our classifiers. In addition, we set $\alpha$ in Equation~\ref{eq:alpha} to be 0.5 and 0.8 for events {\it Moscow airport bombing} and {\it Egyptian revolution} respectively, as the later involves many sub-events we want to capture. 

\noindent{\it (7) Other parameters.} As the number of topical words-related instances in each chunk is less than 2000 on average, we choose to set small numbers on the parameters in Procedure~\ref{alg:criteria}. By teasing on different values on these parameters, we finally settled on selecting around 10\% of the instances as candidate instances, among which only 10\% of the instances are chosen as representative instances, in order to avoid topic drifting.

\begin{table}[t]
\scriptsize
{
\caption{Companion words in event {\it Moscow airport bombing}}
\label{tab:words}
\centering
\begin{tabular}{|c|l|}
\hline
 \textbf{ChunkID} & \textbf{Companion words} \\ \hline
Jan-24	 & \textbf{injured, killed, blast, explosion, 31, dead, bombing, suicide}      \\ \hline
Jan-25	 & \textbf{blast, news, 35, killed, terrorist, attack, bombing, suicide, dead} \\ \hline
Jan-26	 & \textbf{modern, revenge, russian}, heathrow, girlfriend, video, \textbf{bombing}  \\ \hline
Jan-27	 & san, \textbf{russian}, news, \textbf{bombing}, call, \textbf{police}, ap, sings  \\ \hline
Jan-28	 & international, san, gatwick, blvd, \textbf{domodedovo},  \textbf{terror},  \textbf{attack},  \textbf{bombing}  \\ \hline
Jan-29	 & malaga, \textbf{blast}, shut, \textbf{investigators}, orlando, latest, \textbf{bombing}  \\ \hline
Jan-30	 & san, international, passengers, stream, watch, \textbf{security,} people, live  \\ \hline
Jan-31	 & san, london, news, subject, \textbf{bomber}, introduce, luton, \#egypt  \\ \hline
Feb-02	 & source, \textbf{russian}, \textbf{victim}, dfw, international  \\ \hline
Feb-03	 &  cairns, open, townsville, richmond, international, opening, {\bf security}  \\ \hline
Feb-07	 &  \textbf{islamist}, san, \textbf{umarov, ordered}, guardian, \textbf{doku}, operators, \textbf{rebel} \\ \hline
Feb-08	 &  \textbf{claims, umarov, ordered, leader, doku, rebel, bombing, chechen} \\ \hline
\end{tabular}
}
\end{table}

\subsection{Performance of Relevant Content Filter}
To demonstrate the performance of {\it relevant content filter} of WS$^2$FS, we focus on its fundamentals, i.e., {\it companion words}, and its performance on the fixed testing dataset. In Table~\ref{tab:words}, we showed the companion words in event {\it Moscow airport bombing} where the words in bold are event-relevant ones. We can find that at the early stage of the framework, the companion words are closely related to this event. As the topic/event starts to die out or submerge by new topics, the selected companion words starts to slightly drift from the main theme. However, when there are updated information, the given topic/event comes alive again. Overall, the companion words selected by {\it relevant content filter}  can capture the evolution of the events to a large extent.

\begin{figure}[t]
\centering
\includegraphics[width=1.0\linewidth]{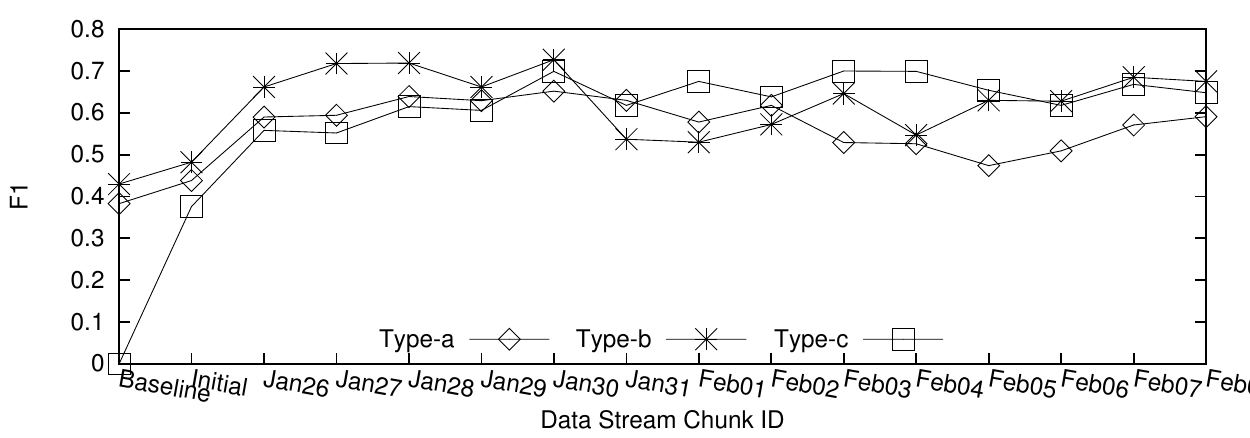}
\caption{Performance of {\it relevant content filter} on event {\it Egyptian revolution} with different types of {\it weak supervision}.}
\label{fig:egyptian}
\end{figure}

In Figure~\ref{fig:egyptian}, we show the F1 score of {\it content relevant filter} on the testing dataset of event {\it Egyptian revolution}, with three types of {\it weak supervision} defined earlier. As we can see, based on {\it Type-a weak supervision}, it retrieves the relevant content quite well, mainly because general concept is universally acknowledged and widely used to discuss about this event. However, the performance begins to decrease as the event evolves, which is probably because people tend to discuss more detailed aspects of the event along its evolution. Armed with the hashtag based {\it Type-b weak supervision}, the {\it relevant content filter} filters out the relevant content with high accuracy in the first few chunks. Later on, as the event evolves, more dynamic ``labels'' are created to describe the event, so the performance simply based on the hashtag slightly decreases in some chunks. Overall, the trending topic/ hashtag born with the event can capture the main theme of the event along with its lifetime. As shown in this figure, the F1 score generated by \emph{Baseline} method with {\it Type-c weak supervision} is around 0. This is because it simply classifies all the tweets containing any of those keywords as relevant content. But later on, our {\it content relevant filter} produces much more accurate relevant content as the event evolves. In general, the {\it relevant content filter} of W$S^{2}$FS can effectively capture the dynamically changing relevant content of an event, with different levels of {\it weak supervision}.

\subsection{Performance of Summarizer}
\subsubsection {Comparison summarization methods}
To compare the {\it summarizer} of WS$^2$FS, we choose the following commonly used classical summarization methods as baseline methods: \\
(1) {\it Centroid:} the centroid instance in the dataset is chosen to be the candidate summary instance~\cite{radev2004centroid}. \\
(2) {\it LexRank:} each instance is modeled as a vertex and the edges are created based on the cosine similarity of the TF-IDF vectors of the two vertices. Graph ranking method (e.g. PageRank) is used to select candidate summary instances~\cite{erkan2004lexrank}.\\
(3) {\it Query-based method:} from the perspective of information retrieval, the summary instance is chose from the most relevant documents to the given query.  It contains the following three different kinds of similarity measures:
\begin{itemize}
\item {\it QueryCosine (Q1)}: uses cosine similarity of the TF-IDF vectors. 
\item {\it  QueryCosineNoIDF (Q2)}: uses cosine similarity on TF vectors. 
\item {\it QueryWordOverlap (Q3)}: uses the overlapping of uni-grams in both document and query to measure the similarity. 
\end{itemize}

\subsubsection {Evaluation criteria for generated summaries}
We compare these baseline methods with the {\it summarizer} of WS$^2$FS on both {\it chunk-wise summary} and {\it sequential summary}. To guarantee the quality of the gold-standard summaries, we ask the annotators to grasp the ``big moments'' along the timeline of the events. 
Specifically, we extract some facts happened on different dates about the event {\it Moscow airport bombing}, based on its Wikipedia page\footnote{\scriptsize{\url{http://en.wikipedia.org/wiki/Domodedovo_International_Airport_bombing}}}.
For event {\it Egyptian revolution}, the annotations should capture the key facts follow the timelines provided by both its Wikipedia page\footnote{\scriptsize{\url{http://en.wikipedia.org/wiki/Timeline_of_the_Egyptian_Revolution_of_2011}}} 
and the report from Al Jazeera English\footnote{\scriptsize{\url{http://www.aljazeera.com/news/middleeast/2011/01/201112515334871490.html}}}. 
The two annotators manually choose the top-3 (if available) most important tweets in each stream chunk as the gold-standard {\it chunk-wise summary}. Accordingly, using the selection criterion {\it diversity} described in Section~\ref{sec:summarization}, the top-3 tweets with highest score generated by {\it summarizer} of WS$^2$FS and all the baseline methods are regarded as their chunk-wise summaries. Each of the two annotators also manually creates a sequential summary along with all the chunks. Using the same selection criterion {\it diversity}, the {\it summarizer} of WS$^2$FS and each of the baseline methods also generate their own sequential summaries for both events. 

\begin{table}[t]
{
\caption{Quality of {\it chunk-wise summary} on event {\it Moscow airport bombing} (average value).}
\label{tab:singleChunk}
\centering
\begin{tabular}{|c||c|c|c|c|c|c|}
\hline
& {\it Centroid} & {\it LexRank} & {\it Q1} & {\it Q2}  & {\it Q3}& {\it WS$^{2}$FS}  \\ \hline \hline
Precision & 0.176  & 0.247  &  0.518   &  0.607  &  0.586   & \textbf{0.706} \\ \hline
Recall & 0.235 & 0.167 & 0.385 & 0.374  & 0.316  & \textbf{0.726} \\ \hline
F1 &  0.200 & 0.195 & 0.434 & 0.449 & 0.394 & \textbf{0.714} \\ \hline
 \end{tabular}
}
\end{table}

\begin{table}[t]
\scriptsize
{
\caption{Quality of {\it chunk-wise summary} on event {\it Egyptian revolution} (average value).}
\label{tab:Chunk_3types}
\centering
\begin{tabular}{|c||c|c|c|c|c|c|}
\hline
& {\it Centroid} & {\it LexRank} & {\it Q1} & {\it Q2}  & {\it Q3}& {\it WS$^{2}$FS}  \\ \hline \hline
Precision & 0.272(b) & 0.232(c)   &  0.192(c)    &  0.203(c)  &  0.205(c)   & \textbf{0.597}(b) \\ \hline
Recall & 0.277(b) & 0.287(c) & 0.359(b) & 0.274(c)   & 0.336(b) & \textbf{0.576}(b) \\ \hline
F1 &  0.277(b) & 0.252(c) & 0.242(c) & 0.229(c)  & 0.245(c) & \textbf{0.585}(b) \\ \hline
 \end{tabular}
}
\end{table}

\begin{figure}[t]
\centering
\includegraphics[width=1.0\linewidth]{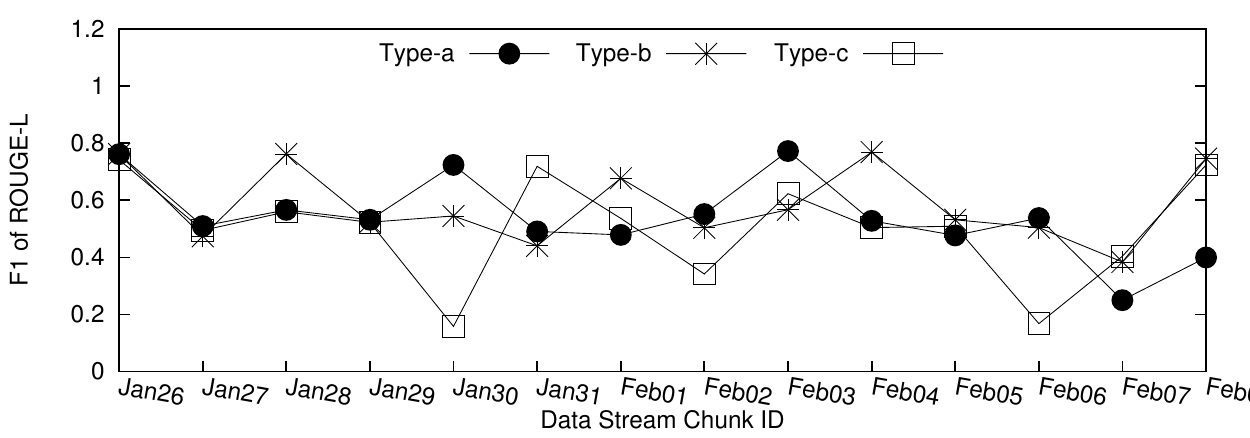}
\caption{Quality of {\it chunk-wise summary} on event {\it Egyptian revolution} in each chunk, produced by {\it summarizer} of W$S^{2}FS$ with different types of {\it weak supervision}.}\label{fig:chunkEvent2}
\end{figure}

\subsubsection {Quality of chunk-wise summary}
In Table~\ref{tab:singleChunk}, we show the quality of {\it chunk-wise summary} generated by different summarization methods on event {\it Moscow airport bombing}. Due to the space limitations, we only list the average values across all chunks. As we can see, our summarizer of WS$^{2}$FS produces the best results in terms of {\it precision}, {\it recall} and {\it F1}. That is, the chunk-wise summary generated by WS$^{2}$FS are most similar to the manually generated summary.

In Table~\ref{tab:Chunk_3types}, we demonstrate the best average ROUGE-L scores of chunk-wise summarization produced by corresponding {\it weak supervision} type (indicated in the parenthesis) on event {\it Egyptian revolution}. On the whole, our summarizer produces the best results using {\it Type-b} supervision. For the baseline methods, the best results are usually generated using {\it Type-b} and {\it Type-c weak supervision}. This is because they largely rely on topical words, and can not fully explore the general concept ({\it Type-a}).

Furthermore, we are interested in investigating how different types of {\it weak supervision} affect the performance of our {\it summarizer} in producing summary in each chunk along with the event evolution. The F-1 values on each chunk are shown in Figure~\ref{fig:chunkEvent2}. In general, the performance fluctuates more with {\it Type-c weak supervision}, probably due to the matching ``degree'' between its keywords and the corresponding big moments of the event. In the real scenario, what keywords will be associated to an unscheduled event is nearly unable to predict. Thus, the method which can generate good summary based on hashtag ({\it Type-b}) or general information ({\it Type-a}) is more useful than the method which rely on specific topical words ({\it Type-c}), which again verified the practical usefulness of our proposed framework which makes better use of {\it Type-b} supervision.

\begin{figure*}[t]
        \centering
        \begin{subfigure}[b]{0.32\textwidth}
                \includegraphics[width=\textwidth]{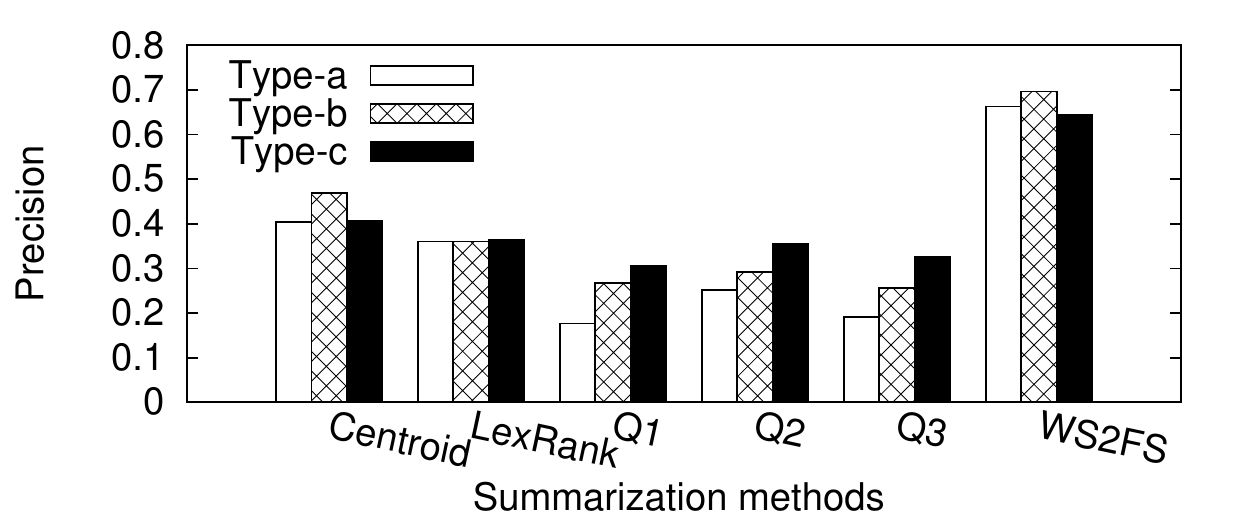}
                \caption{Precision}
        \end{subfigure}
        \begin{subfigure}[b]{0.32\textwidth}
                \includegraphics[width=\textwidth]{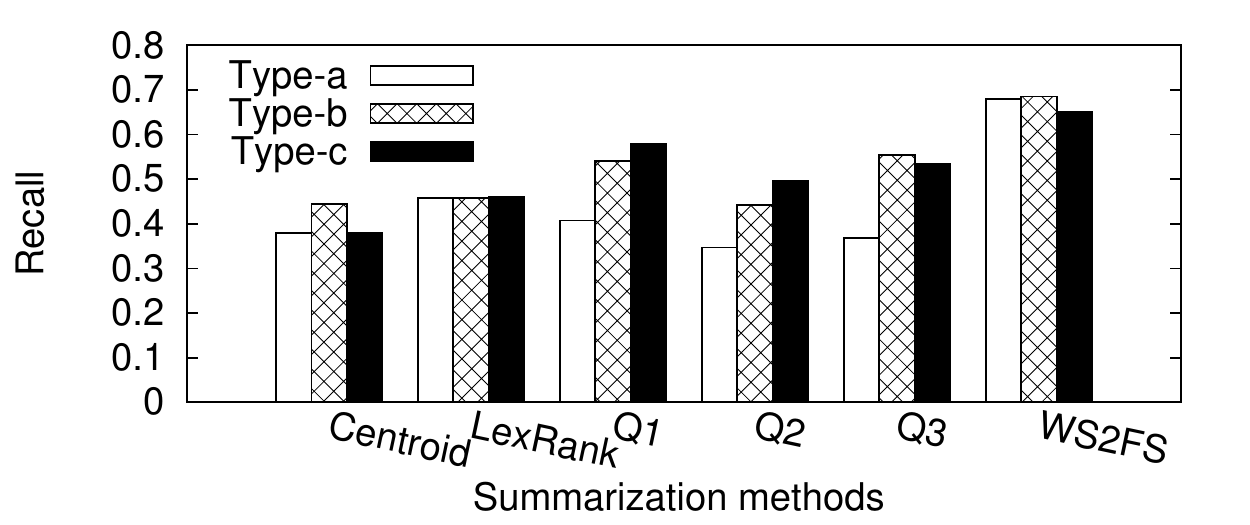}
                \caption{Recall}
        \end{subfigure}
        \begin{subfigure}[b]{0.33\textwidth}
                \includegraphics[width=\textwidth]{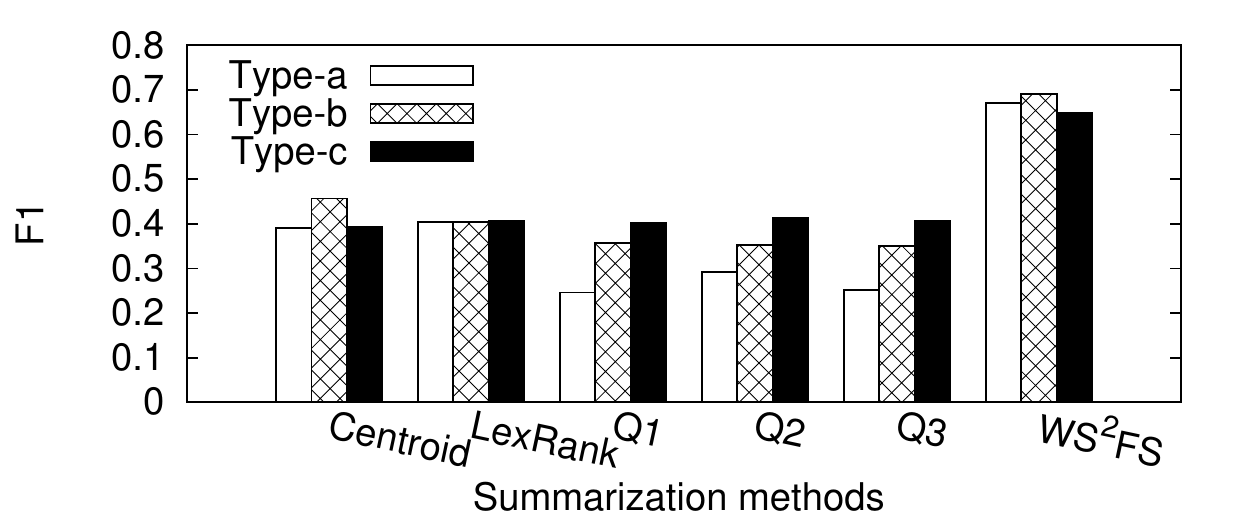}
                \caption{F1}
        \end{subfigure}
        \caption{Quality of {\it sequential summary} on event {\it Egyptian revolution} with different types of {\it weak supervision}.}
        \label{fig:sequential}
\end{figure*}

\subsubsection {Results of sequential summarization}
Table~\ref{tab:sequential} shows the quality of sequential summaries generated by different methods on event {\it Moscow airport bombing}, compared with the the gold-standard sequential summary. From this table, we can see that {\it query-based} methods tend to produce higher precision but lower recall. {\it Centroid} and {\it LexRank} produce poor results overall. Our {\it summarizer} produces the best results. The evaluation of sequential summarization on event {\it Egyptian revolution} are shows in Figure~\ref{fig:sequential}, with different methods under different types of {\it weak supervision}. It shows the similar results as of the chunk-wise summaries in Table~\ref{tab:Chunk_3types}. That is, W$S^2$FS produces the best results. In terms of {\it weak supervision}, those baseline methods benefit more from {\it Type-c weak supervision} as they rely more on topical words. W$S^2$FS produces the best results with hashtag-based {\it Type-b weak supervision}, and generates good enough results based on {\it weak supervision} coming from general concept ({\it Type-a}) as well. The resultant sequential summary produced by the {\it summarizer} on the two events are publicly available along with the gold standard summaries\footnotemark[1] .

\begin{table}[t]
{
\caption{Quality of {\it sequential summary} on event {\it Moscow airport bombing}.}
\label{tab:sequential}
\centering
\begin{tabular}{|c|c|c|c|c|c|c|}
\hline
& {\it Centroid} & {\it LexRank} & {\it Q1} & {\it Q2}  & {\it Q3}& {\it WS$^{2}$FS}  \\ \hline \hline
Precision & 0.267 & 	0.439 & 	0.544 & 	\textbf{0.682} & 	0.649  & 0.631 \\  \hline
Recall & 0.338 & 	0.177 & 	0.416 & 	0.457 & 	0.374 & 	\textbf{0.662} \\ \hline
F1 & 0.299 & 	0.252 & 	0.471 & 	0.547 & 	0.474 & 	\textbf{0.646} \\  \hline
 \end{tabular}
 }
\end{table}

\section{Discussion}
\label{sec:discussion}
One of the important inputs to our system is {\it weak supervision}, this weak supervision should not only be easy to obtain for a variety of events but also be effective in generating meaningful summaries. In our experiments, we have shown that it is possible to achieve both goals simultaneously. More specifically, we have experimented and shown results for both {\it relevant content filtering} and {\it summarization} based on three different types of {\it weak supervision} that are relatively easier to obtain. 
Among these three types of weak supervision, {\it type-b} weak supervision (i.e. trending-topic/hashtag based) has performed the best. This hashtag-based weak supervision also happen to be the one that can be most easily obtained. Other than its easy availability, hashtag-based weak supervision also has an immediate practical advantage. For the users who tweet with a particular hashtag, the proposed summarization method can be used to provide them an up-to-date summary of the events related to their hashtags, which is a very useful feature to be integrated in social media. In addition to the personalized event summary, the framework can also be used to provide global event summary based on the hashtags trending on the site. Such advantage demonstrates the practicality of our method and strengthens the argument for its adaptation into real world applications. While our method is able to exploit such power of trending topics, unfortunately those baseline methods cannot do so. If one were to treat the initial weak supervision as a query, then our method is able to produce the dynamic query-based summary (i.e. evolution of the event) as opposed to the static summaries produced by the baselines. It is important to note that our method relies on trending topics only for initial weak supervision, and updates itself as the new data comes in, unlike other query based summarization methods which entirely depend on the initial query to produce summaries.

\section{Conclusion}
\label{sec:conclusion}

In this paper, we have presented a relevant content filtering based framework for data stream summarization in social media platforms. This framework does not use any labeled data as typical supervised methods do; it instead uses the weak supervision in the forms of rules and guidelines. Such weak supervision makes this framework extensible for different applications. In addition, our framework is built for streaming environment, i.e. it is not only able to filter out the relevant content with regard to the given topic or event, but also can generate the sequential summary or event story line as it evolves in the text stream. Our experiments on a set of tweets about two real events verified its effectiveness in generating summarization from data stream after relevant content filtering. Besides, our experimental results showed that using the most easy-to-obtain weak supervision, i.e., hashtags, it can generate the best results. Such property makes our framework more applicable: it can be regarded as an almost unsupervised framework given the trending topic or hashtag; it can be easily used to generate both personal event summary and global event summary, based on the personal defined hashtags and trending hashtags representing global events, respectively. Thus, our framework can be easily integrated into social media platforms such as Twitter.

\bibliographystyle{IEEEtran}
\bibliography{IEEEabrv,MicroblogSummarization}

\end{document}